# Images Within Images? A Multi-Image Paradigm with Novel Key-Value Graph Oriented Steganography

Subhrangshu Adhikary

Department of Computer Science and Engineering, Dr. B.C. Roy Engineering College, Durgapur, West Bengal, India – 713206
`subhrangshu.adhikary@spiraldevs.com`

**Abstract.** Steganographic methods have been in the limelight of research and development for concealing secret data within a cover media without being noticed through general visualization. The Least Significant Bits (LSBs) of 8-bit color code for the RGB image arises the possibility of replacing the last two bits with the bits of the encrypted message. Several procedures have been developed to hide an image within another image however in most cases the payload image has to be within the accommodatable range of the cover image and very little literature have shown methods to hide multiple images within multiple images. This paper presents a novel approach to split the image into JSON styled dictionary of key-value pairs and using a metadata graph to locate different parts and positions of the payload images in the entire cluster of cover images. The model could be easily used in the real world scenario for privately sharing secret data over public communication channels without being noticed.

## 1 Introduction

The quest for developing methods to ensure privacy for data transmission over public communication channels has given rise to different steganographic techniques [1]. Steganography is the method of hiding data in a host file without being noticed and is particularly used in environments where directly applying cryptographic encryption arise suspicion [2]. Based on the host medium, different types of steganography include hiding data in images, sounds, text, video or other computer files. Images are one of the most widely used steganographic cover media [3]. Different methods can be applied for hiding data within images and these are generally Least Significant Bit (LSB) Insertion, Masking and Filtering techniques, Redundant Pattern Encoding, Encrypt and Scatter and finally Coding and Cosine transformation [4]. Among all these, LSB Insertion is the simplest and most widely used technique [5]. This method does not invoke suspicion as a significant amount of data can be hidden within an image and the size of the stego image is always close to the original image [6]. In this method, stego image is created by replacing the least significant bits of the image with the secret data. Using 1 bit generally alters the pixel color intensity by approximately

±1 unit and therefore the stego image is almost indistinguishable however this gives a very little capacity for storing data [7]. Using 3 bits alters the pixel color intensity of up to ±5 units which gives a large capacity to store data but the stego image has visible differences compared to the original image [8]. Therefore generally 2 bits are used to store the data which alters the color intensity by up to ±3 units and therefore the differences between the stego image and original image have very little differences and are indistinguishable by human eyes and besides this, using 2 bits gives a decent capacity to store the data [9].

JavaScript Object Notation (JSON) is a format of data communication based on dictionaries of key-value pair where '{' and '}' are used to start and end a block, keys are surrounded by ".." quotation marks and values are written based on their corresponding data types [10]. The data can be nested and values can be of array types as well. This method can be used to store metadata of the dataset as well as different values associated with the intermediate stages of the secret data [11].

Most steganographic works for hiding data within images based on LSB has been performed to hide text data however recently different techniques of flattening the image matrix have become popular to store the image data within an image, however, when the payload file size is large, the cover media fails to accommodate the secret image [12-13]. Other techniques have evolved to accommodate data of a single image over multiple cover images but very lesser studies have been performed to store multiple secret images spread across multiple cover images [14-15]. This motivated us to develop this novel method to store multiple secret images across multiple cover images by flatmapping the image, splitting it into chunks and arranging JSON with metadata and graph based mapping. The details of the technique and performances are discussed later in the text.

## 2  Methodology

The proposed methodology for hiding and retrieving of multiple images within multiple images having undergone multiple intermediate steps. The details have been explained in the following text.

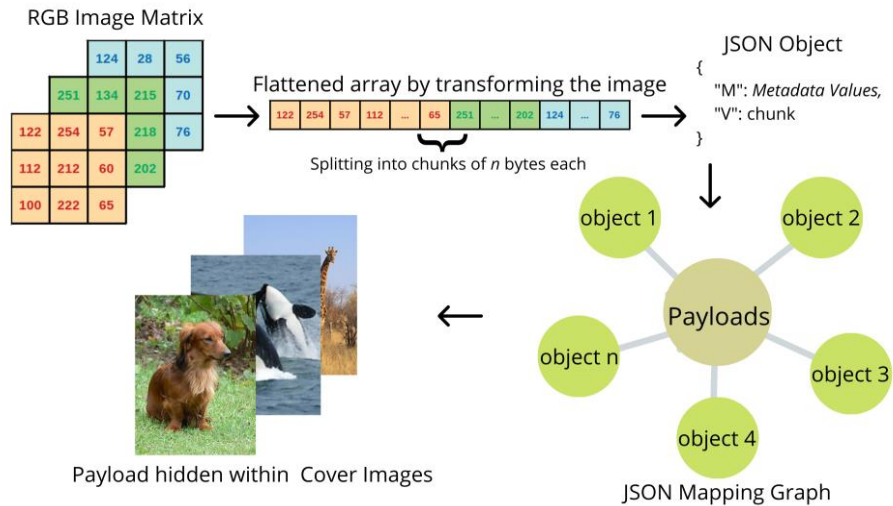

**Fig. 1.** The representation of the proposed multi-image paradigm method where the RGB matrix of the image is first flattened and dividing into chunks of *n* bytes which along with the metadata properties of the image are then used to convert into JSON Objects. Following this, the objects are used to create the JSON Mapping graph in the form of a string and the string is finally embedded into the cover image with LSB insertion method

### 2.1 Processing the payload images

The most important part of the proposed algorithm is processing the payload images to embed them in such a way that it is easily recovered even after distributing the data in multiple images [16]. For this purpose, the flatmap technique has been used to convert a single image matrix into a 1D array. In this, the matrix for each of the colors namely Red, Greed and Blue have been individually converted into a 1D array by appending each row one after another and finally appending a 1D array for each color one after another. The same process is repeated for each of the payload images.

After this step, each of these 1D arrays has been split into chunks of size based on the cover size of cover images. If the chunk size is too small, a lot of data is consumed by JSON Graph metadata but very little amount of pixels are left vacant in this process. When the chunk size is too large, the size of JSON Graph metadata is small however a large number of pixels in the cover image are left vacant. Therefore keeping these in mind, 512 byte chunks have been used for the experiment.

### 2.2 The mapping graph and JSON key-value pair dictionary

After flattening the payload images and dividing them into chunks of suitable sizes, the mapping graph is created using JSON. For this purpose, the first two key-value pair of the JSON contains the number of payload and cover images. Followed by this, the child of the nested JSON objects contain details of each image [17-18]. In these, the metadata contains the shape of the original payload image including height, width, colormaps, number of chunks, etc. And the corresponding image-data key contains the array of tuples containing the position of the chunk in the graph, numbering to identify the payload image and the processed chunks of 512 bytes each. This JSON is then used to prepare the final string to be embedded within the cover images [19-20].

### 2.3 LSB Insertion to the cover images

Followed by the processing of the payload images and preparing the JSON graph string for each payload image, the prepared payload strings are then inserted into the least two significant bits of each of the image pixels [21-22]. Each bit of the payload strings is serially incorporated within the cover images one after another based on their encodable capacity according to the general Steganographic LSB insertion method. Hence the stego images are generated [23-24].

### 2.4 Retrieval of payload images

Once the stego images have been generated, it is also required to be decoded to retrieve the payload. For this, the LSBs of all the stego images are individually decoded and combined to the required location based on the graph mapping within the JSON strings of each stego image. Then the combined string is arranged to merge the split chunks for each image separately and then based on the metadata for the images, the flatmap 1D array is reshaped to form the image matrix and finally, the hidden images are retrieved [25-26].

## 3 Results & Discussion

The LSB insertion of the payload string was prepared by combining JSON objects of chunks of flattened RGB matrix forming the mapping graph. The text has been performed on 8 cover images and 2 payload images. The 8 cover images include photographs of a dog, whale, giraffe, horse, squirrel, camel, tiger and fish and the 2 payload images includes a photograph of a cat and a parrot as shown in figure 2. Combining these 8 images, a total of 38115048 slots of 2 bits each are the available space to store data while using 2 LSB for each point. The 2 images of the payload combined require space of 35047586 slots of 2 bit however because of the creation of the JSON Mapping Graph and padding, approximately an additional 5% space is required making the payload string size 37228972. This makes utilization of 97.67% available space. On a computer of Intel i3 $6^{th}$ Gen CPU and 12 Gigabytes of RAM the method required 74.7 seconds for encoding and 19.1 seconds for decoding the images.

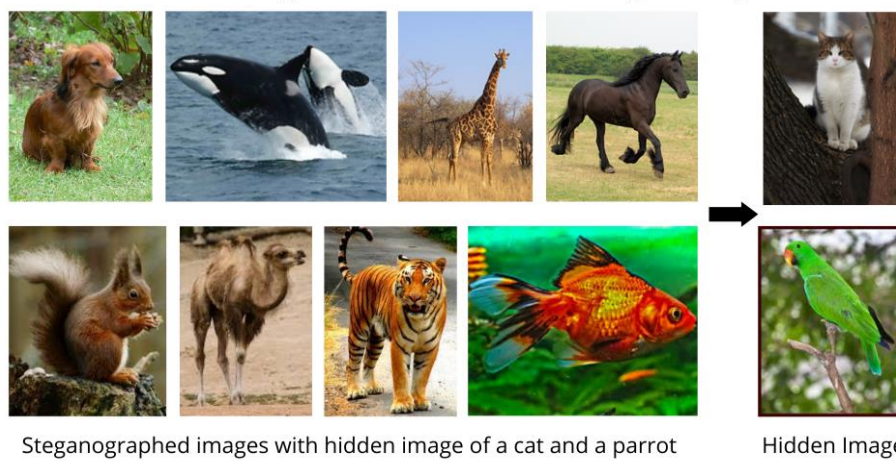

**Fig. 2.** The figure demonstrates the 8 stego images on the left which had hidden the two payload images in the right without being visibly noticed

The stego images are visually indifferent from the original image. However, the differences are clear on observing their color intensity histogram. The histogram for one image, for example, whale compared to its steganographed counterpart have been shown in figure 3. It can be observed that although there are indetectable visual differences between the original image and stego image, their histograms have significant differences denoting the modification of the LSBs. The maxima of the counts for specific intensity values of all pixels combined for the original image was around 110000 however the maxima for the stego image was around 175000. Observing the color bands individually, it could be noticed that the red color have maximum occurrence near 110 intensity values for both original and stego image however the maximum number of times the red value appeared is around 78000 original image and around 123000 for stego image. Similarly for green, the intensity maxima had occurred for both the images at around 140 but the highest number of times the intensity value has occurred is around 76000 for the original image and around 125000 for stego image. Finally, the blue color maxima had occurred around the intensity value range near 165 for both the images but the maxima for the original image is approximately around 80000 and for stego image is around 124000.

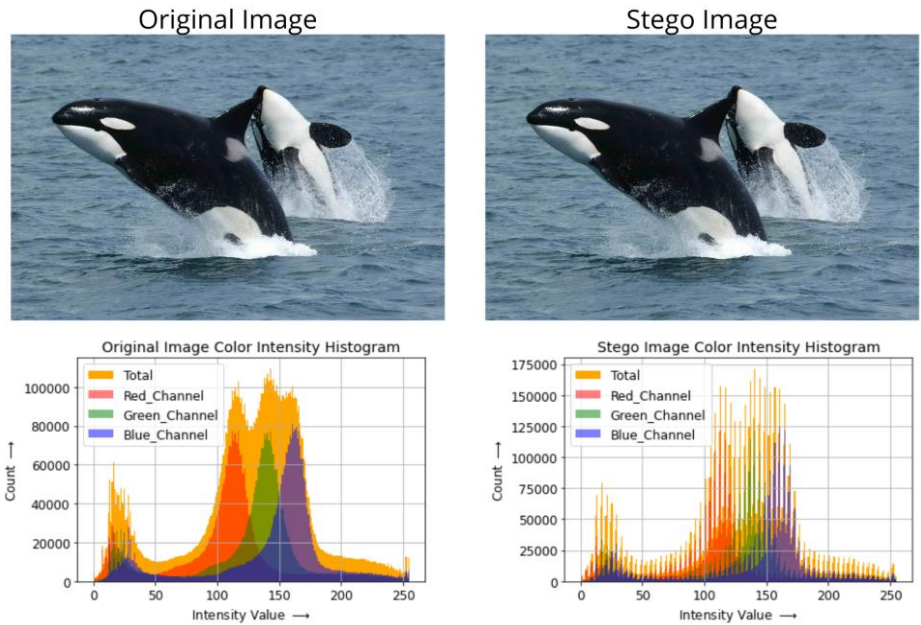

**Fig. 3.** The side by side comparison of original image and stego image with their corresponding histogram of color intensity

At a closer look, it can be observed that the histogram of the original image has followed an evenly distributed pattern without sudden spikes but the histogram of stego image appears discrete with gaps at almost regular intervals. This is because, in a natural image, the neighbouring pixel values change smoothly besides sharp boundaries of different colors, textures or intensities. On the other hand, the last two bits of the stego image has been altered causing a difference of ±3 intensities for every neighbouring pixel as well as for all three color matrix. This breaks the natural continuity of the changing color and makes the histogram discrete. This is a limitation of the LSB method as well where although the differences of the images cannot be distinguished visually but a simple histogram can be easily used to detect whether an image is steganographed and this is why encrypting the payload with cryptographic techniques are also recommended along with steganography to completely safeguard the data. This phenomenon also explains why the maxima of the color intensity for all the colors is much higher for stego image compared to the original image as the gaps within the color intensity values have been appended to the spikes of the histogram of the stego image.

## 4 Conclusion

The steganographic methods are used to hide messages within another cover media without being noticed. The Least Significant Bit (LSB) insertion method is a popular steganographic method to hide data within an image making use of very insignificant bits. Different approaches have been used to hide an image within another image however they have a limitation of encodable capacity. This paper presents a method to solve this issue by introducing a novel approach with JSON Mapping Graph based to encode multiple images within multiple images.

The method uses flatmap technique on RGB image matrix and then split the data into chunks of *n* bytes each and then JSON object is created with those chunks and metadata of the image. Finally, the objects hence created are used to create the Mapping Graph which finally creates the encodable payload string. These strings are then embedded into the cover images by LSB insertion method. The method is visually indistinguishable but can be detected with histogram and hence cryptographic encryption is also suggested on the payload string for added safety.

The method can be easily used to embedded large payload image files over multiple cover images in any order and then can be combined to get back the payload file. The stego images can be shared over public communication channels without being visually noticed maintaining privacy. Further, the model could be improved to fit very large data files within images without being noticed and reduce the discrete histogram spikes with more evenly spread spikes.

**Acknowledgments.** The work is a part of The Gyanam Project as a joint collaboration between Spiraldevs Automation Industries Pvt. Ltd., India and CubicX, India.